\pdfoutput=1

\documentclass[11pt]{article}

\usepackage[]{acl}

\usepackage{times}
\usepackage{latexsym}

\usepackage[T1]{fontenc}

\usepackage[utf8]{inputenc}

\usepackage{microtype}

\usepackage{inconsolata}

\usepackage{graphicx}

\usepackage{booktabs}
\usepackage{multirow}
\usepackage{amsmath, amssymb}
\usepackage{array}
\usepackage{tcolorbox}
\usepackage{subcaption}

\title{Unmasking Fake Careers: Detecting Machine-Generated Career Trajectories via Multi-layer Heterogeneous Graphs}

\author{
  Michiharu Yamashita$^{1}$ \quad 
  Thanh Tran$^{2}$ \quad 
  Delvin Ce Zhang$^{3}$ \quad 
  Dongwon Lee$^{1}$ \\
  $^{1}$The Pennsylvania State University \quad
  $^{2}$Amazon \quad
  $^{3}$University of Sheffield \\
  \texttt{\{michiharu, dongwon\}@psu.edu}, 
  \texttt{tdt@amazon.com}, 
  \texttt{delvin.ce.zhang@sheffield.ac.uk}
}

\newcommand{\ours}{{\textsf{CareerScape}}}

\newcommand{\ie}{{\textit i.e.}}
\newcommand{\eg}{{\textit e.g.}}

\begin{document}
\maketitle
\begin{abstract}

The rapid advancement of Large Language Models (LLMs) has enabled the generation of highly realistic synthetic data. We identify a new vulnerability, LLMs generating convincing {\em career trajectories} in fake resumes and explore effective detection methods. 
To address this challenge, we construct a dataset of machine-generated career trajectories using LLMs and various methods, and demonstrate that conventional text-based detectors perform poorly on structured career data. We propose {\ours}, a novel heterogeneous, hierarchical multi-layer graph framework that models career entities and their relations in a unified global graph built from genuine resumes. Unlike conventional classifiers that treat each instance independently, {\ours} employs a structure-aware framework that augments user-specific subgraphs with trusted neighborhood information from a global graph, enabling the model to capture both global structural patterns and local inconsistencies indicative of synthetic career paths. Experimental results show that {\ours} outperforms state-of-the-art baselines by 5.8-85.0\% relatively, highlighting the importance of structure-aware detection for machine-generated content. Our codebase is available at \url{https://github.com/mickeymst/careerscape}.
\end{abstract}

\section{Introduction}
\begin{figure*}[th]
\centering
\includegraphics[width=\linewidth]{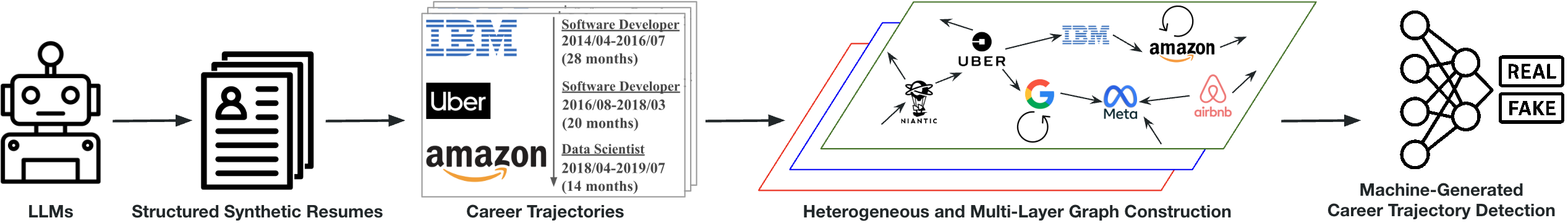}
\caption{Motivation and overview of machine-generated career trajectory detection.}
\label{fig:attack}
\end{figure*}

Large Language Models (LLMs) have revolutionized text generation, enabling the production of highly realistic content across unstructured general texts and structured formats such as JSON, CSV, and tables~\cite{li2024pre, liu2024llms, vijayan2023prompt, elnashar2025enhancing}. While these advancements drive progress in multiple fields \cite{tang2023does, kumichev2024medsyn, lee2025large, moore2023empowering}, they also cause novel security threats \cite{das2025security}. One emerging threat is the automatic generation of fake resumes, which can infiltrate recruitment pipelines, manipulate screening through data poisoning, and erode trust in job platforms \cite{yamashita2024fake}. 


Real-world incidents highlight this severity: over 1,000 non-existent SpaceX engineers were identified on LinkedIn\footnote{\url{http://tiny.cc/etri001}}. With LLMs' growing accessibility, malicious actors can easily fabricate plausible career histories and mass-produce fake accounts for fraudulent purposes. An April 2025 report shows that ``scammers leverage generative AI to fabricate resumes, photo IDs, and employment histories,'' threatening both job seekers and employers\footnote{\url{http://tiny.cc/ktri001}}. 
Beyond inflating user numbers, fabricated career paths can distort recommendation systems, reduce job visibility for genuine candidates, and destabilize labor markets. Such profiles are also exploited in financial scams, directly harming job seekers and employers. 

Prior research on machine-generated text detection has mainly focused on linguistic features, text representations, and coherence-based analysis  \cite{jawahar2020automatic, sadasivan2023can, mitchell2023detectgpt, macko2023multitude, lucas2023fighting, venkatraman2024gpt, solaiman2019release}. While effective for traditional textual domains such as news articles and online reviews, these methods fall short in settings where text can be organized into structured representations, such as resumes, where information is encoded through entities. Unlike generic text, resumes contain inherent structured elements, such as career trajectories and job roles, which can be systematically represented as structured data \cite{qin2023comprehensive}. When LLM-generated resumes are converted to structured formats, conventional detectors lose effectiveness due to the loss of linguistic artifacts, presenting a unique detection challenge. 

To address this challenge, we reframe machine-generated career trajectory detection as a structured fraud detection problem and propose {\ours}, a heterogeneous and hierarchical multi-layered graph neural network framework that captures complex relationships and entity-level dynamics in career data. 
In our approach, each user's career history is represented as a graph of job titles, companies, and descriptions connected via meaningful relations. By exploiting both intra-layer and cross-layer relationships, {\ours} identifies structural anomalies and semantic inconsistencies unique to synthetic career paths. 
Importantly, {\ours} introduces a novel structure-based augmentation paradigm for career data. We construct a global heterogeneous graph from genuine resumes, integrating career entities into a unified structural space. Rather than using this global graph for representation learning, we utilize its topology to expand each local graph into a \textit{global-augmented subgraph} by incorporating trusted neighboring nodes and edges. This context-aware approach allows each career graph with realistic structural context, enabling {\ours} to expose subtle inconsistencies in machine-generated trajectories that deviate from real-world patterns, such as implausible title transitions or company sequences. Such structural augmentation is especially effective in the career domain, where transitions tend to follow semantically coherent and domain-informed patterns. 

To this end, we construct a comprehensive dataset of machine-generated resumes using multiple generative strategies, and our experiments show that {\ours} consistently outperforms state-of-the-art text-based and graph-based detectors. Our key contributions are: 
1) We formalize \textbf{machine-generated career trajectory detection as a structured fraud detection task}, revealing limitations of conventional text-only detectors; 
2) We propose {\ours}, \textbf{a heterogeneous and hierarchical multi-layered graph framework} that models career entities and their relations within a global structure built from genuine career trajectories;
3) We introduce a novel \textbf{global-to-local subgraph augmentation approach} that expands user-specific subgraphs from a global graph, enabling detection of subtle structural inconsistencies in synthetic career paths; 
4) We demonstrate the effectiveness of {\ours} through extensive experiments and ablation studies, contributing to more secure and trustworthy online job platforms. 

\section{Related Work}
\noindent {\bf Machine-Generated Text Detection.} 
Detection of machine-generated text (MGT) has gained prominence with advancing Large Language Models (LLMs) \cite{crothers2023machine}. Approaches evolved from perplexity-based methods identifying fluency discrepancies \cite{abburi2023simple, tang2024science} to neural classifiers \cite{solaiman2019release}, coherence modeling \cite{wu2024detectrl}, external knowledge verification \cite{wang2024ideate}, and psycholinguistic features \cite{venkatraman2024gpt}. While effective for unstructured text, these methods target linguistic patterns rather than structural inconsistencies, limiting their efficacy in domains where text can be transformed into entity-based representations, such as resumes. 

\noindent {\bf Machine-Generated Profiles.}
AI-generated content poses significant security threats across various domains, including online job platforms \cite{das2025security}, with recent studies showing fake resumes can manipulate job recommendation systems \cite{yamashita2024fake}. While previous work has explored fake profile detection using account metadata, social signals, or textual features \cite{chakraborty2022fake, roy2021fake, xiao2015detecting, yuan2019detecting}, our work focuses specifically on career trajectories-central to professional platforms such as LinkedIn. We argue that LLM-generated plausible career paths require structure-aware detection methods beyond existing detectors. Graph structures have proven effective in career trajectory modeling \cite{yamashita2022looking, lee2024caper}, highlighting the need for detection methods that capture entity-level and trajectory-level inconsistencies. 

\noindent {\bf Graph-level Classification.}
Graph-based methods have been widely applied in fraud detection by leveraging relational structures. Heterogeneous graph neural networks \cite{zhang2019heterogeneous} have been applied to financial fraud \cite{wu2024heterogeneous} and fake news detection \cite{ren2020adversarial, zeng2022heterogeneous}. While prior graph-based fraud detection approaches operate on isolated graphs, overlooking globally shared structures, our method constructs a unified heterogeneous graph spanning all resume entities. This enables global-augmented subgraphs across users, effectively capturing both structural and semantic inconsistencies in machine-generated career trajectories.

\section{Dataset Generation}\label{sec:dataset}

To effectively train and evaluate our machine-generated resume detection framework, we construct a dataset of both real and machine-generated resumes in a structured format centered on career trajectories. Inspired by real-world online job platforms such as LinkedIn, we focus on core components typically emphasized in user profiles: job titles, companies, and employment durations, motivating our dataset design. We provide an example of a LinkedIn profile to illustrate a typical career section in Appendix \ref{sec:linkedin_example}. 

\subsection{Real Resume Data}

We curate real resumes from two datasets used in career trajectory research \cite{yamashita2024fake}, representing technology and business domains. Each resume is parsed into structured entities including job titles, companies, and employment durations. To ensure industry-wide generalizability, we combine both domains. For graph construction, we retain resumes containing companies that appear more than three times, resulting in a dataset of 4,555 verified genuine resumes. Expanding to additional domains (\eg, healthcare and education) remains an important direction for future work. 

\subsection{Threat Model} 

Our synthetic resume generation assumes the following threat model: 1) Malicious actors create fake accounts with convincing career trajectories for whatever purposes they intend (\eg, facilitate scams and identity theft); 2) Creating multiple accounts on professional platforms is easy and inexpensive; 3) For credibility, attackers associate trajectories with legitimate companies and create plausible job titles and progression; 4) Attackers have access to modern LLMs and sufficient prompt engineering knowledge. This threat model reflects real-world challenges where detecting synthetic identities represents a critical security concern. Our synthetic resume generation methods are designed to simulate these adversarial capabilities. 

\subsection{Synthetic Resume Construction}\label{sec:generators}

We employ diverse generation methods to create synthetic resumes, producing 1,000 examples for each category, ensuring balanced representation across generation strategies. 

\noindent {\bf Rule-Based Generators}. 
These methods manipulate real resume data using predefined heuristics. We implement four strategies: \textit{``Random''} combines randomly selected entities, creating syntactically valid but unrealistic progressions; \textit{``Popular''} assembles resumes using top ten percent entities with log-normal distributed job durations \cite{meng2019hierarchical, yamashita2024openresume}; \textit{``Swapping''} randomly exchanges companies within real resumes; and \textit{``Replacing''} substitutes companies with alternatives from other resumes. 

\noindent {\bf LLM-Based Generators}. 
We utilize the following LLM-based generators in both zero-shot and few-shot settings: \textit{GPT-4o}~\cite{hurst2024gpt}, \textit{LLaMA-3}~\cite{grattafiori2024llama}, and \textit{Gemini-2.0}~\cite{team2023gemini}. The exact prompts are detailed in Appendix \ref{sec:prompt_settings}. 

\noindent {\bf Agent-Based Generators}. 
We employ an agent-based generator that autonomously refines resumes through iterative self-improvement. Prompts and agent configurations are detailed in Appendix \ref{sec:prompt_settings}. 

\noindent {\bf Domain-Specific Generators}. 
We also incorporate a state-of-the-art domain-specific generator, FRANCIS \cite{yamashita2024fake}, which is a data poisoning framework that generates realistic resumes to manipulate job recommendation systems. FRANCIS optimizes resume content to evade detection, providing an ideal benchmark for evaluating robust fake resume detection methods. 


\subsection{Job Description Data}
To obtain consistent and informative job descriptions for each resume entry, we leverage JAMES \cite{yamashita2023james} to normalize job titles into standardized ESCO categories\footnote{\url{https://esco.ec.europa.eu/en}}. We then retrieve corresponding descriptions from the official ESCO job title description dataset. This establishes a one-to-one mapping between normalized titles and descriptions, enhancing semantic richness and enabling graph-based models to reason over standardized content.

\subsection{Assessing LLM-Generated Resumes} 
Assessing qualities of the generated resumes with human evaluators poses a few challenges: 
1) showing real resumes to human evaluators is not permitted, as individuals can potentially be re-identified from their unique career trajectories, and 
2) the use of the original real dataset is governed by an MOU that prohibits sharing the data with individuals (\ie, human evaluators) not covered under the agreement. 
To ensure compliance with IRB policies, 
we use LLM-based quality assessment instead of human evaluation and
perform a realism evaluation using Claude-3.7, a strong LLM not used in our generation step nor main experiments.

\vspace{0.05in}
\noindent{\bf Setup.} 
We randomly sample 100 resumes each from GPT-4o, LLaMA-3, Gemini-2.0, Agent, and our real dataset. We ask Claude-3.7 to assess realism on a {\em 1–5 scale}, where 1 indicates clearly artificial content and 5 indicates highly realistic, human-authored content. The prompt (Appendix \ref{sec:real_asses}) avoids revealing the source and requests impartial judgment.

\begin{figure}[tb]
    \centering
    \includegraphics[width=0.75\linewidth]{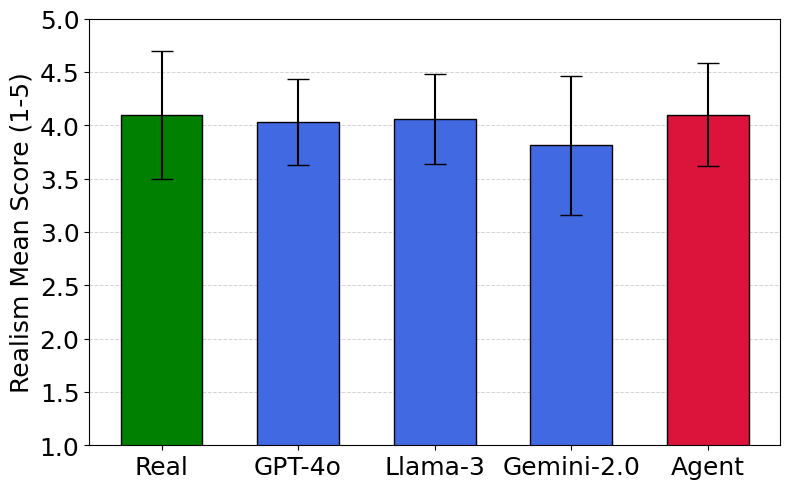}
    \caption{LLM-generated resume reality assessment.}
    \label{fig:realism}
\end{figure}

\vspace{0.05in}
\noindent{\bf Results.} 
LLM-generated resume realism scores are comparable to real resumes, with average differences below 0.3 points. \textit{GPT-4o} and \textit{LLaMA-3} achieve the highest scores, not statistically different from real resumes. Figure~\ref{fig:realism} presents average scores with standard deviations. These findings suggest LLM-generated resumes demonstrate fluency and realism on par with human-written ones, reinforcing the validity of using these resumes as surrogates and supporting their inclusion in our generated dataset.

\section{Problem Definition}\label{sec:problem}

\begin{table}[t]
\centering
\scriptsize
\caption{Notation used throughout the paper.}
\begin{tabular}{p{0.25\linewidth} p{0.63\linewidth}}
\toprule
\textbf{Notation} & \textbf{Description} \\
\midrule
\( \mathcal{G}_u = (\mathcal{V}_u, \mathcal{E}_u) \) & Structured subgraph representing the career trajectory of user \( u \) \\
\( \mathcal{V}_t, \mathcal{V}_c, \mathcal{V}_d \) & Sets of job title, company, and job description nodes \\
\( \mathcal{V}_u \subseteq \mathcal{V} \) & Node set in user subgraph \( \mathcal{G}_u \) \\
\( \mathcal{E}_u \subseteq \mathcal{E} \) & Edge set in user subgraph \( \mathcal{G}_u \) \\
\( \phi: \mathcal{V} \rightarrow \mathcal{T}_v \) & Node-type mapping function (\eg, job title, company, description) \\
\( \psi: \mathcal{E} \rightarrow \mathcal{T}_e \) & Edge-type mapping function (\eg, worked\_at, has\_description) \\
\( \mathcal{N}_r(v) \) & Neighbors of node \( v \) under relation type \( r \) \\
\( \mathbf{h}_v^{(l)} \in \mathbb{R}^d \) & Embedding vector of node \( v \) at GNN layer \( l \) \\
\( \mathbf{d}_{uv} \in \mathbb{R}^{d_d} \) & Embedding for job duration on edge \( (u,v) \) \\
\( W_r^{(l)} \in \mathbb{R}^{d \times d} \) & Relation-specific transformation matrix at layer \( l \) \\
\( \tilde{\mathbf{h}}_v \in \mathbb{R}^d \) & Node embedding with type embedding added before encoder \\
\( \mathbf{z}_u \in \mathbb{R}^d \) & Aggregated representation of user subgraph \( \mathcal{G}_u \) \\
\( \hat{y}_u \in [0, 1] \) & Predicted probability that \( \mathcal{G}_u \) is machine-generated \\
\( y_u \in \{0,1\} \) & Ground-truth label: 1 for synthetic, 0 for human-authored \\
\( \mathcal{L} \) & Binary cross-entropy loss function for training \\
\bottomrule
\end{tabular}
\label{tab:notations}
\end{table}



Our goal is to develop a detection model that distinguishes between genuine and machine-generated resumes. Let \( \mathcal{G}_u = (\mathcal{V}_u, \mathcal{E}_u) \) denote a structured subgraph representing user \( u \)'s career trajectory. The node set \( \mathcal{V}_u \) includes entities such as job titles, companies, and job descriptions, connected through semantic and temporal relations that capture both flow and context of the career path. 

Given this representation, the task is to determine whether a given career trajectory is machine-generated. 
Formally, we aim to learn a function \( f_\theta \) that maps the input subgraph \( \mathcal{G}_u \) to a probability score \( \hat{y}_u \in [0, 1] \), where the target label \( y_u \in \{0,1\} \) indicates whether the career trajectory is synthetic (1) or human-authored (0):
\( \hat{y}_u = f_\theta(\mathcal{G}_u). \)
See Table \ref{tab:notations} for a summary of the main mathematical notations. We also provide task motivation and challenges of text-based detection in Appendix \ref{sec:challenge_of_text}.



\section{The Proposed Model}\label{sec:proposed}

\begin{figure*}[t]
    \centering
    \includegraphics[width=\linewidth]{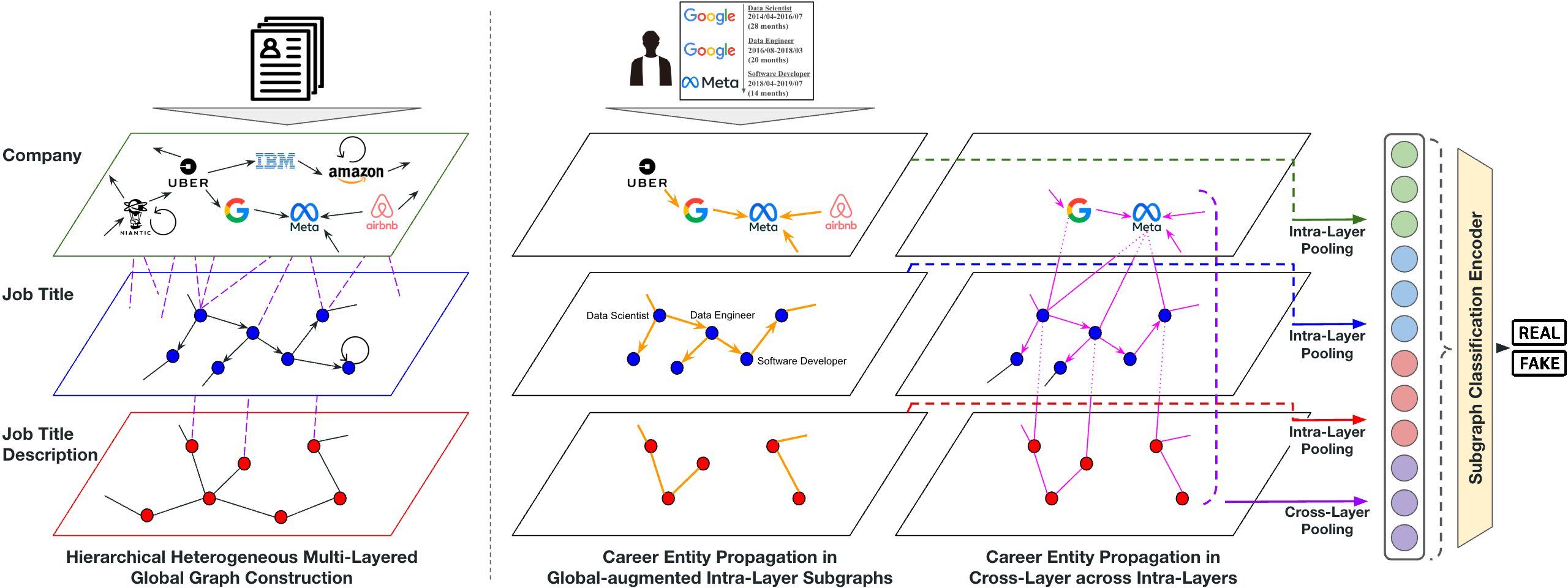}
    \caption{Our proposed model, {\ours}'s architecture.}
    \label{fig:model}
\end{figure*}

{\ours} is designed to leverage the structured and relational characteristics inherent in career data by representing each user's career history as a heterogeneous multi-layer graph composed of job titles, companies, and job descriptions. 
Unlike conventional graph classifiers that treat each instance as an isolated graph, {\ours} introduces a structure-aware, global-to-local modeling paradigm. We first construct a unified global graph using real resumes, capturing reliable structural patterns among career entities. During classification, each user-specific subgraph is expanded into a \textit{global-augmented subgraph} by incorporating trustworthy neighboring nodes and edges from the global graph. This augmentation introduces realistic contextual structure that helps identify subtle anomalies in machine-generated trajectories. 
See Figure \ref{fig:model} for the overview of {\ours}.

\subsection{Key Novelty of {\ours}}
{\ours} introduces the following key novelty distinguishing from prior work: 
1) \textbf{Novel task formulation}: We address a new task setting where each local graph shares common node entities but exhibits unique structural patterns, enabling construction of a global graph from real instances and requiring structure-aware subgraph classification; 
2) \textbf{Global-to-local subgraph augmentation}: Instead of treating each graph independently, {\ours} augments each subgraph with contextual structure derived from a global graph constructed solely from real resumes, enabling structure-aware reasoning; 
3) \textbf{Multi-layer heterogeneous graph representation}: We model career trajectories using a rich heterogeneous graph that spans job titles, companies, and descriptions, capturing both intra-type and inter-type entity relations. 
%
%

While prior work has focused on subgraph classification in homogeneous or locally sampled settings \cite{zhang2018link, khoshraftar2024survey}, few studies have addressed subgraph-level classification within a globally constructed heterogeneous graph. Moreover, existing heterogeneous GNNs typically focus on node- or edge-level tasks (\eg, classification, prediction, recommendation). 

{\ours} is the first to introduce structure-driven subgraph augmentation over a global constructed heterogeneous graph, bridging local reasoning with global structural context in a novel task setting. By leveraging this design, each resume can incorporate complementary information from others through the global graph, thereby compensating for missing details and enabling more accurate detection.




\subsection{Heterogeneous Career Entities}
We define a heterogeneous graph \( \mathcal{G} = (\mathcal{V}, \mathcal{E}, \phi, \psi) \), where \( \mathcal{V} \) is the set of nodes, \( \mathcal{E} \subseteq \mathcal{V} \times \mathcal{V} \) is the set of edges, \( \phi: \mathcal{V} \rightarrow \mathcal{T}_v \) maps each node to its type (job title, company, or job description), and \( \psi: \mathcal{E} \rightarrow \mathcal{T}_e \) maps each edge to a relation type.

The node set is composed of \( \mathcal{V} = \mathcal{V}_t \cup \mathcal{V}_c \cup \mathcal{V}_d \), where \( \mathcal{V}_t \), \( \mathcal{V}_c \), and \( \mathcal{V}_d \) denote job title, company, and job description nodes, respectively.

Each node \( v \in \mathcal{V} \) is associated with an embedding \( \mathbf{h}_v^{(0)} \in \mathbb{R}^{d} \): job titles and companies are initialized as learnable embeddings, while job descriptions are initialized using pre-trained JobBERT \cite{zhang2022skillspan} embeddings followed by a linear projection. Job durations, being continuous values, are encoded as real-valued scalars and mapped to \( \mathbb{R}^{d_d} \) via a linear layer and incorporated into edge representations.

\subsection{Multi-Layer Heterogeneous Graphs} 
The graph is hierarchically organized with both intra-layer and cross-layer connections. The intra-layer graphs include: \textit{(i) Job Title Graph} \( \mathcal{G}_t \), where edges represent job transitions; \textit{(ii) Company Graph} \( \mathcal{G}_c \), capturing transitions between companies; and \textit{(iii) Job Description Graph} \( \mathcal{G}_d \), constructed using cosine similarity over JobBERT embeddings,
\( \mathcal{E}_d = \{(u, v) \mid \cos(\mathbf{h}_u, \mathbf{h}_v) \geq \tau\}, \)
where \(\tau\) is a similarity threshold hyperparameter empirically set to 0.9. 
The cross-layer graphs include: \textit{(i) Worked\_at edges} between job titles and companies, and \textit{(ii) Has\_description edges} from job titles to job descriptions.

\subsection{Model Architecture and Learning}

\noindent {\bf Heterogeneous Graph Embedding}. 
Each node \( v \) is initialized with \( \mathbf{h}_v^{(0)} \). Our learning process involves both intra-layer and cross-layer message passing to capture the hierarchical structure of career trajectories. 
For intra-layer message passing, nodes aggregate information from neighbors of the same type via relation-specific transformations. For example, job titles learn from chronologically adjacent job titles, capturing career progression patterns. 
For cross-layer message passing, nodes exchange information across different entity types. Job titles gather context from associated companies and descriptions, while companies learn from their connections to various job roles. 
This dual-level message passing is formalized at layer \( l+1 \) as:
\[
\mathbf{h}_v^{(l+1)} = \sigma \left( \sum_{r \in \mathcal{T}_e} \sum_{u \in \mathcal{N}_r(v)} \frac{1}{|\mathcal{N}_r(v)|} W_r^{(l)} \mathbf{h}_u^{(l)} \right),
\]
where \( \mathcal{N}_r(v) \) is the set of neighbors via relation type \( r \), \( W_r^{(l)} \) is the relation-specific weight matrix learned during training, and \( \sigma \) is ReLU activation \cite{agarap2018deep}. This allows the model to jointly learn from both intra-layer patterns and cross-layer relationships. 
Job duration information is particularly important for career trajectory analysis. For \textit{worked\_at} relations, we incorporate duration embeddings \( \mathbf{d}_{uv} \) into edge-level messages:
\(
\mathbf{m}_{uv}^{(l)} = W_r^{(l)} \left[ \mathbf{h}_u^{(l)} \, \| \, \mathbf{d}_{uv} \right],
\)
where \( \| \) denotes concatenation. This enables the model to distinguish between short-term positions and long-term roles when reasoning about career patterns. 

After \( L \) layers of passing, the final node embeddings \( \mathbf{h}_v^{(L)} \) capture both local career dynamics and global structural patterns, incorporating information from all entity types and their relationships.

\noindent {\bf Subgraph Augmentation and Representation}. 
A key aspect of {\ours} is using the global graph \( \mathcal{G}_{global} \) solely for topological augmentation. Prior to message passing, each user-specific subgraph \( \mathcal{G}_u \) is expanded using neighboring nodes from the global graph, enabling the model to reason over contextualized structures that reflect realistic career patterns. 
Specifically, we augment each user subgraph \( \mathcal{G}_u \) by including first-order neighbors of its nodes within a predefined hop threshold in the global graph. This expansion adds contextually relevant entities, such as co-occurring job titles or semantically related companies observed in real trajectories, without learning on the global graph itself. 
Message passing is then performed only on the augmented subgraph, allowing the model to aggregate signals from both the original career path and surrounding entities that provide realistic structural context.

For each augmented subgraph \( \mathcal{G}_u \), we collect the final-layer embeddings \( \{\mathbf{h}_v^{(L)}\}_{v \in \mathcal{V}_u} \). Type-specific embeddings \( \mathbf{t}_v \) are added to each node embedding:
\(
\tilde{\mathbf{h}}_v = \mathbf{h}_v^{(L)} + \mathbf{t}_v.
\) 
The resulting sequence \( \tilde{\mathbf{H}}_u \) is processed by a self-attention-based subgraph encoder:
\(
\mathbf{H}'_u = \text{SubgraphEncoder}(\tilde{\mathbf{H}}_u).
\) 
We apply global mean pooling to obtain the subgraph-level representation:
\(
\mathbf{z}_u = \frac{1}{|\mathcal{V}_u|} \sum_{v \in \mathcal{V}_u} \mathbf{H}'_u[v].
\) 
Finally, a linear classification head with sigmoid activation predicts the likelihood of being machine-generated:
\[
\hat{y}_u = \sigma(\mathbf{w}^\top \mathbf{z}_u + b), \quad \hat{y}_u \in [0, 1].
\]

\subsection{Model Training and Optimization Details}
We optimize {\ours} using a binary classification objective that distinguishes machine-generated career trajectories from genuine ones. Instead of applying an MLP to a pooled embedding, we adopt a self-attention-based classifier that takes node embeddings within the augmented user-specific subgraph as input, enabling richer modeling of inter-node dependencies.  
The training objective minimizes the binary cross-entropy loss:
\[
\mathcal{L} = -\frac{1}{N} \sum_{i=1}^{N} \left[ y_i \log(\hat{y}_i) + (1 - y_i) \log(1 - \hat{y}_i) \right],
\]
where \( y_i \in \{0,1\} \) is the ground-truth label and \( \hat{y}_i \) is the predicted probability. 
Training uses Adam optimizer \cite{kingma2014adam} with learning rate scheduling and dropout for regularization. For robustness to unseen entities, a single universal (UNK) embedding is used for unknown companies. 
We use 128 as embedding dimension, 0.005 as initial learning rate in the subgraph encoder. For augmentation, we set the hop threshold to 2, and apply dropout with rate 0.1. The model is trained for 100 epochs with early stopping (patience=10). 


\section{Experiments}\label{sec:exp} 

\begin{table*}[t]
\centering
\small
\caption{F1-Score for machine-generated career trajectory detection. We conducted {$t$}-tests at a 95\% confidence level. The superscript {$^*$} indicates statistically significant differences between our model and competitors ($p < 0.05$).}
\resizebox{\textwidth}{!}{
\begin{tabular}{ll||cc|cc|cc|ccc|c}
\toprule
\multicolumn{2}{c||}{} & \multicolumn{10}{c}{\textbf{\color{red}{Detector}}} \\
\cmidrule{3-12}
\multicolumn{2}{c||}{} & \multicolumn{2}{c|}{\color{red}{ML-based}} & \multicolumn{2}{c|}{\color{red}{LLM-based (GPT-4o)}} & \multicolumn{2}{c|}{\color{red}{MGT detectors}} & \multicolumn{3}{c|}{\color{red}{Graph-based}} & \color{red}{Multi-layer} \\
\cmidrule{3-12}
\multicolumn{2}{l||}{\textbf{\color{blue}{Generator}}} & \textbf{XGBoost} & \textbf{LightGBM} & \textbf{Zero-shot} & \textbf{Few-shot} & \textbf{GPTZero} & \textbf{DetectGPT} & \textbf{HAN} & \textbf{GraphSAGE} & \textbf{R-GNN} & \textbf{{\ours}} \\
\midrule
\multirow{4}{*}{\textcolor{blue}{Rule-based}} & Random & 0.62 & 0.60 & 0.58 & 0.64 & 0.53 & 0.55 & 0.72 & 0.67 & 0.70 & \textbf{0.78$^*$} \\
& Popular & 0.61 & 0.62 & 0.59 & 0.65 & 0.65 & 0.56 & 0.85 & 0.94 & 0.94 & \textbf{0.96$^*$} \\
& Swapping & 0.62 & 0.62 & 0.57 & 0.57 & 0.58 & 0.56 & 0.61 & 0.64 & 0.65 & \textbf{0.69$^*$} \\
& Replacing & 0.64 & 0.63 & 0.57 & 0.63 & 0.56 & 0.56 & 0.68 & \textbf{0.80} & 0.76 & \textbf{0.80\ \ } \\
\midrule
\multirow{3}{*}{\textcolor{blue}{LLM-based}} & GPT-4o & 0.67 & 0.67 & 0.41 & 0.52 & 0.46 & 0.39 & 0.75 & 0.82 & 0.81 & \textbf{0.85$^*$} \\
& LLaMA-3 & 0.70 & 0.69 & 0.40 & 0.52 & 0.40 & 0.41 & 0.78 & 0.81 & 0.80 & \textbf{0.84$^*$} \\
& Gemini-2.0 & 0.70 & 0.68 & 0.41 & 0.63 & 0.52 & 0.40 & 0.81 & 0.80 & 0.79 & \textbf{0.84$^*$} \\
\midrule
\textcolor{blue}{Agent-based} & Agent & 0.62 & 0.63 & 0.39 & 0.43 & 0.62 & 0.38 & 0.75 & 0.78 & 0.80 & \textbf{0.82$^*$} \\
\midrule
\textcolor{blue}{Domain-specific} & FRANCIS & 0.60 & 0.60 & 0.56 & 0.62 & 0.51 & 0.46 & 0.57 & 0.61 & 0.61 & \textbf{0.69$^*$} \\
\midrule
\textcolor{blue}{Combined} & Combined & 0.60 & 0.62 & 0.51 & 0.61 & 0.54 & 0.49 & 0.78 & 0.80 & 0.80 & \textbf{0.86$^*$} \\
\bottomrule
\end{tabular}
}
\label{tab:fake_resume_results_f1}
\end{table*}

\subsection{Evaluation Setup} 
We perform our evaluation on the dataset described in Section~\ref{sec:dataset}, splitting it into 80\% for training (with 20\% of that reserved for internal validation) and 20\% for testing, where the global graph is constructed exclusively from training data and never includes test data, preventing data leakage. Experiments are performed separately for each generator type, as well as on a ``Combined'' dataset, which includes a balanced sample comprising 10\% from each generator type along with real resumes. To the best of our knowledge, there is no public dataset for detecting machine-generated career trajectories. As a result, all experiments are carried out on our constructed dataset. 

\subsection{Baseline Detectors}\label{sec:baseline}
We compare against four categories of baselines: 

\noindent {\bf ML-Based Detectors}. 
We implement \textit{XGBoost} \cite{chen2016xgboost} and \textit{LightGBM} \cite{ke2017lightgbm} as feature-based state-of-the-art machine learning baselines.

\noindent {\bf LLM-Based Detectors}. 
Leveraging LLMs' self-detectability \cite{lucas2023fighting}, we use \textit{GPT-4o} as a detector in both zero-shot and few-shot settings, converting structured career trajectories to text using templates provided in Appendix \ref{sec:llm_detector}. 

\noindent {\bf MGT Detectors}. 
We employ \textit{GPTZero}\footnote{\url{https://gptzero.me/}} and \textit{DetectGPT} \cite{mitchell2023detectgpt} as state-of-the-art MGT detectors via text templates (Appendix \ref{sec:template}).

\noindent {\bf Graph-Based Detectors}. 
While our detection task via graphs itself is new with no existing dedicated models, we adopt \textit{HAN} \cite{wang2019heterogeneous}, \textit{GraphSAGE} \cite{hamilton2017inductive}, and \textit{R-GNN} \cite{wu2020comprehensive} as compatible graph-level classification baselines. 

\subsection{Result and Key Findings}

\begin{figure*}[t!]
    \centering
    \begin{subfigure}[b]{0.327\linewidth}
        \includegraphics[width=\linewidth]{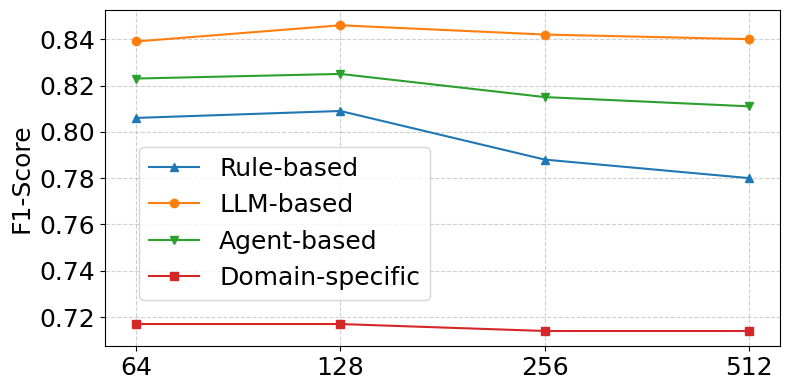}
        \caption{Embedding Size}
        \label{fig:embsize}
    \end{subfigure}
    \hfill
    \begin{subfigure}[b]{0.327\linewidth}
        \includegraphics[width=\linewidth]{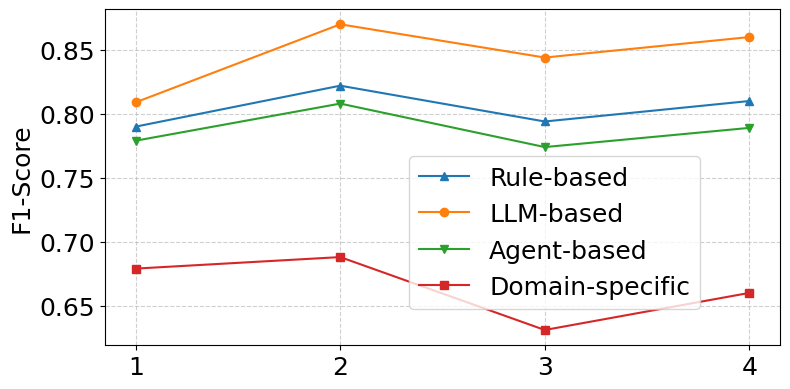}
        \caption{Propagation Depth (Hops)}
        \label{fig:hops}
    \end{subfigure}
    \hfill
    \begin{subfigure}[b]{0.327\linewidth}
        \includegraphics[width=\linewidth]{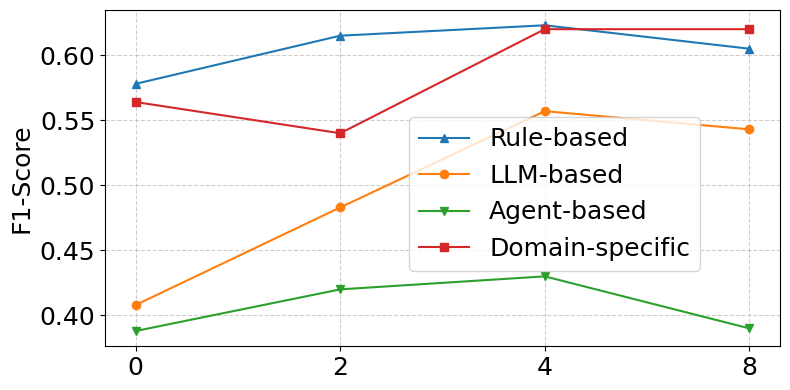}
        \caption{Few-shot Sample Count of GPT-4o}
        \label{fig:shots}
    \end{subfigure}

    \caption{Impact of model configurations on detection performance.}
    \label{fig:ablation_config}
\end{figure*}

Table \ref{tab:fake_resume_results_f1} shows micro F1-scores across different generator types. {\ours} significantly outperforms other baseline methods in most categories ($p < 0.05$). 
Our experiments reveal several key insights: 
1) \textbf{Superiority of structure-aware detection:} 
{\ours} achieves the highest performance (F1-score 0.86 on the combined dataset), demonstrating the effectiveness of heterogeneous multi-layered graph representations in capturing structural inconsistencies of fake trajectories. 
2) \textbf{Limitations of feature-based models:} Traditional ML models (F1-score from 0.60 to 0.70) struggle with sophisticated generators, including domain-specific generators and rule-based generators, which highlights the limitations of purely feature-based classification. 
3) \textbf{Weakness of LLM-based detector:} LLM-based detectors perform poorly on LLM/Agent-generated resumes (F1-score from 0.39 to 0.63), with GPT-4o zero-shot detection showing F1-score as low as from 0.39 to 0.41, revealing a significant blind spot for structured content detection such as career trajectories. 
4) \textbf{Inconsistency in existing MGT detectors:} Existing MGT detectors' performance varies widely (F1-score from 0.38 to 0.65), with poor results on LLaMA-3 and Gemini-2.0 content, indicating overfitting to specific text generation patterns. 
5) \textbf{Benefits of Graph-based models:} Standard graph-based models outperform non-graph approaches. Comparing {\ours} with the graph-based baselines, {\ours} outperforms them due to our novel global-to-local subgraph augmentation. 
6) \textbf{Robustness to challenging generation methods:} {\ours} maintains robust performance against FRANCIS (F1-score 0.69), a domain-specific generator designed to evade detection. Moreover, the swapping-based generation method proved surprisingly effective in creating convincing fake trajectories, with most detectors achieving F1-scores below 0.65. 
The swapping approach poses a notable challenge by preserving nodes and altering only their order, making structural differences subtle and harder to detect. This simple yet effective method may threaten job platforms attempting to identify fake resumes. Despite these challenges, {\ours} outperforms all baselines, highlighting its ability to identify subtle structural anomalies even when significant portions of authentic career patterns are preserved. 

\section{Ablation Study}\label{sec:ablation}
To gain deeper insights into {\ours}'s designs, we conduct comprehensive ablation studies. 

\subsection{Model Configuration Ablation}\label{subsec:model_ablation}
We investigate how embedding size and hops affect detection performance, and analyze GPT-4o's few-shot detection capabilities. 
Figure~\ref{fig:ablation_config} presents average F1 scores across four types of machine-generated datasets for each setting. Figure~\ref{fig:embsize} and \ref{fig:hops} show that moderate configurations yield optimal detection performance, suggesting a trade-off between model complexity and effectiveness. 
Figure~\ref{fig:shots} indicates that increasing shot count improves GPT-4o's few-shot detection, though with varying effectiveness. Rule-based and domain-specific datasets peak at F1-score 0.62 with 4 shots. Interestingly, GPT-4o shows particular weakness in detecting LLM-based and agent-based generation (F1-scores 0.56 and 0.43 respectively with 4 shots), significantly lower than its performance on other types. Further increasing shots would incur higher computational costs without proportional gains, while {\ours} outperforms these few-shot approaches without requiring large prompts.

\subsection{Impact of Structure-Driven Augmentation}

\begin{figure}[t]
    \centering
    \includegraphics[width=\linewidth]{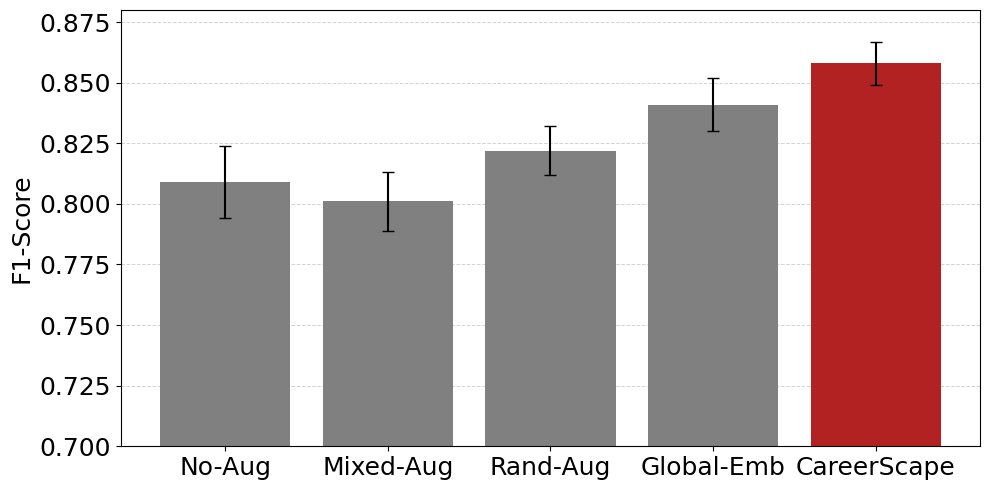}
    \caption{Impact of augmentation.}
    \label{fig:ablation_aug}
\end{figure}


To evaluate structure-driven augmentation's contribution and verify the inductive bias introduced by our global graph, we test several simplified variants on the combined generation dataset and report results in Figure~\ref{fig:ablation_aug}. 
\textbf{No-Aug} removes structural augmentation, using only the original user-specific subgraph.  
\textbf{Rand-Aug} augments each subgraph with randomly sampled nodes from the global graph.  
\textbf{Mixed-Aug} constructs the global graph using both real and fake resumes, introducing noisy connections.  
\textbf{Global-Emb} replaces structure-based augmentation with pretrained global node embeddings. 


Our experiments reveal significant performance differences. {\ours} achieves the highest F1-score 0.86, demonstrating our structure-driven augmentation's effectiveness. {Global-Emb} performs well (F1-score 0.84) but statistically lower than {\ours} ($p < 0.05$), highlighting the advantage of explicit structural augmentation over purely embedding-based representations. 
Removing augmentation ({No-Aug}, F1-score 0.81) or introducing noise ({Mixed-Aug}, F1-score 0.80) substantially degrades performance. Random augmentation ({Rand-Aug}, F1-score 0.82) offers some improvement but still underperforms compared to {\ours}. The clear performance hierarchy ({\ours} > Global-Emb > Rand-Aug > Mixed-Aug $\approx$ No-Aug) confirms that {real-only, structure-driven augmentation provides an effective inductive bias} for detecting machine-generated careers.

\subsection{Graph-layer Ablation}
We further perform a graph-layer ablation to evaluate the contribution of different components in our heterogeneous multi-layer graph architecture, using the combined dataset. 
Table~\ref{tab:model_ablation} shows performance under seven configurations, each removing one or more structural elements of the full model. 
The first group uses only single intra-layer graphs (\ie, job titles, companies, or job descriptions). The second group combines two intra-layers. The third group integrates all three intra-layer graphs, while the final configuration (\textit{All}) includes both intra-layer graphs and cross-layer connections. 
The results demonstrate each graph component contributes meaningfully to the overall performance, with statistically significant differences observed between groups with different numbers of layers ($p < 0.01$). The progression from single-layer to multi-layer configurations shows consistent improvement, with an average increase of 0.13 in F1-score from single-layer to the full model. The inclusion of cross-layer relations and the semantically rich job description graph leads to substantial gains. 
It also shows that job titles are particularly informative for detection, outperforming companies, as unusual role progressions (\eg, CTO$\rightarrow$Sales$\rightarrow$SDE) create more readily detectable anomalies than company movements alone. 
Even the model only using a single layer performs competitively against ML-based and text-based baselines, highlighting the strength of our structured representation.

\begin{table}[t]
\centering
\small
\caption{Impact of different graph layers.}
\label{tab:model_ablation}
\resizebox{\linewidth}{!}{
\begin{tabular}{clc}
\toprule
\textbf{\# of Layers} & \textbf{Graph Layers} & \textbf{F1-Score} \\
\midrule
1 & Job Title (JT) & 0.73 \\
1 & Company (C) & 0.68 \\
1 & Job Description (JD) & 0.72 \\
\midrule
2 & JT + C & 0.75 \\
2 & JT + JD & 0.79 \\
\midrule
3 & JT + C + JD (\ie, all Intra-layers) & 0.83 \\
All & JT + C + JD + Cross-layers (\ie, {\ours}) & \textbf{0.86} \\
\bottomrule
\end{tabular}
}
\vspace{-3pt}
\end{table}



\section{Conclusion}


In this paper, we formalized machine-generated career trajectory detection as a structured fraud detection task and proposed {\ours}, a multi-layer heterogeneous graph framework with global-to-local subgraph augmentation that effectively captures structural patterns and local inconsistencies. Our findings emphasize the importance of structure-aware modeling in detecting synthetic structured data generated by LLMs. 
This work contributes to building more secure online job platforms, with future work focusing on model robustness against adversarially crafted fake resumes.

\section*{Limitations}
While {\ours} achieves strong performance in detecting machine-generated career trajectories, several limitations remain. First, our dataset is limited to the technology and business domains, and future work should expand to other industries for broader applicability and generalizability of our findings. Second, as LLMs continue to advance rapidly, newer generative techniques may produce more sophisticated fake career trajectories, requiring continual updates to detection models. Finally, our approach relies on the availability of structured and sufficiently detailed resume data. In cases where online job platform profiles are extremely sparse or lack key entities (\eg, job titles and companies), the model's ability to capture meaningful patterns may be reduced.

\section*{Ethical Considerations}

This work on detecting machine-generated career trajectories raises several ethical considerations. Our goal aligns with efforts in fake news detection, human-AI content differentiation, and social media bot detection, focusing on identifying fabricated content while respecting authentic diversity. We acknowledge the inherent sensitivity of resume verification and emphasize that our goal is not to flag non-linear or unconventional career paths as fraudulent. 
Our focus is on detecting maliciously fabricated contents that threaten job platform integrity.
All resume data used in this study are anonymized or synthetically generated, with no personally identifiable information involved. Due to the sensitive nature, we have limited our public release to include only LLM-generated datasets. While targeting algorithmically mass-generated fake profiles, we acknowledge potential dual-use concerns that merit careful consideration. We explicitly caution against misappropriation for problematic applications such as employee surveillance or discriminatory candidate filtering, and strongly advocate for implementing frameworks incorporating transparency, fairness, and meaningful human oversight when deploying detection systems within a job platform's security contexts. 
Finally, we call for broader community engagement to ensure fairness, accountability, and the responsible use of AI in employment ecosystems.

\section*{Acknowledgments}
This work was in part supported by NSF awards \#1934782 and \#2131144, and the 2021 seed funding from the Center for Socially Responsible Artificial Intelligence
(CSRAI) at Penn State. 

\bibliography{custom}

\clearpage
\appendix
\section{Motivation and Challenges}\label{sec:challenge_of_text}
\subsection{Task-Specific Motivation}
Career trajectories offer a unique structure-rich domain where individual instances, while diverse, share a high degree of node vocabulary and structural regularity. This makes them especially amenable to structure-based augmentation. Real career paths often exhibit semantically coherent transitions (\eg, promotions, lateral moves within an industry), whereas machine-generated fakes often violate such patterns. By constructing a global graph from real data, {\ours} injects trusted contextual structure into each subgraph, enabling it to better capture these inconsistencies. This design is particularly well-suited for detecting anomalies in career data, where deviations from real-world transitions can be subtle yet critical. 

\subsection{Challenges of Text-Based Detection}
While prior work focuses on detecting AI-generated content through surface-level text features, these methods are insufficient in structured scenarios. Once resume content is parsed and represented as entities and relations (\ie, job titles and companies), linguistic cues become unavailable. Synthetic resumes created by LLMs can mimic realistic entity sequences, making it difficult for text-based detectors to capture deeper inconsistencies in structure or semantics. 

This motivates the need for structure-aware models that can reason over entity-level interactions and the dynamics of career trajectories. Our objective is to build a robust and generalizable detection framework that leverages the heterogeneous and layered nature of structured resume data to uncover subtle artifacts indicative of synthetic generation.

\section{Career Experience Section in LinkedIn}\label{sec:linkedin_example}
\begin{figure}[th]
    \centering    
    \includegraphics[width=\linewidth]{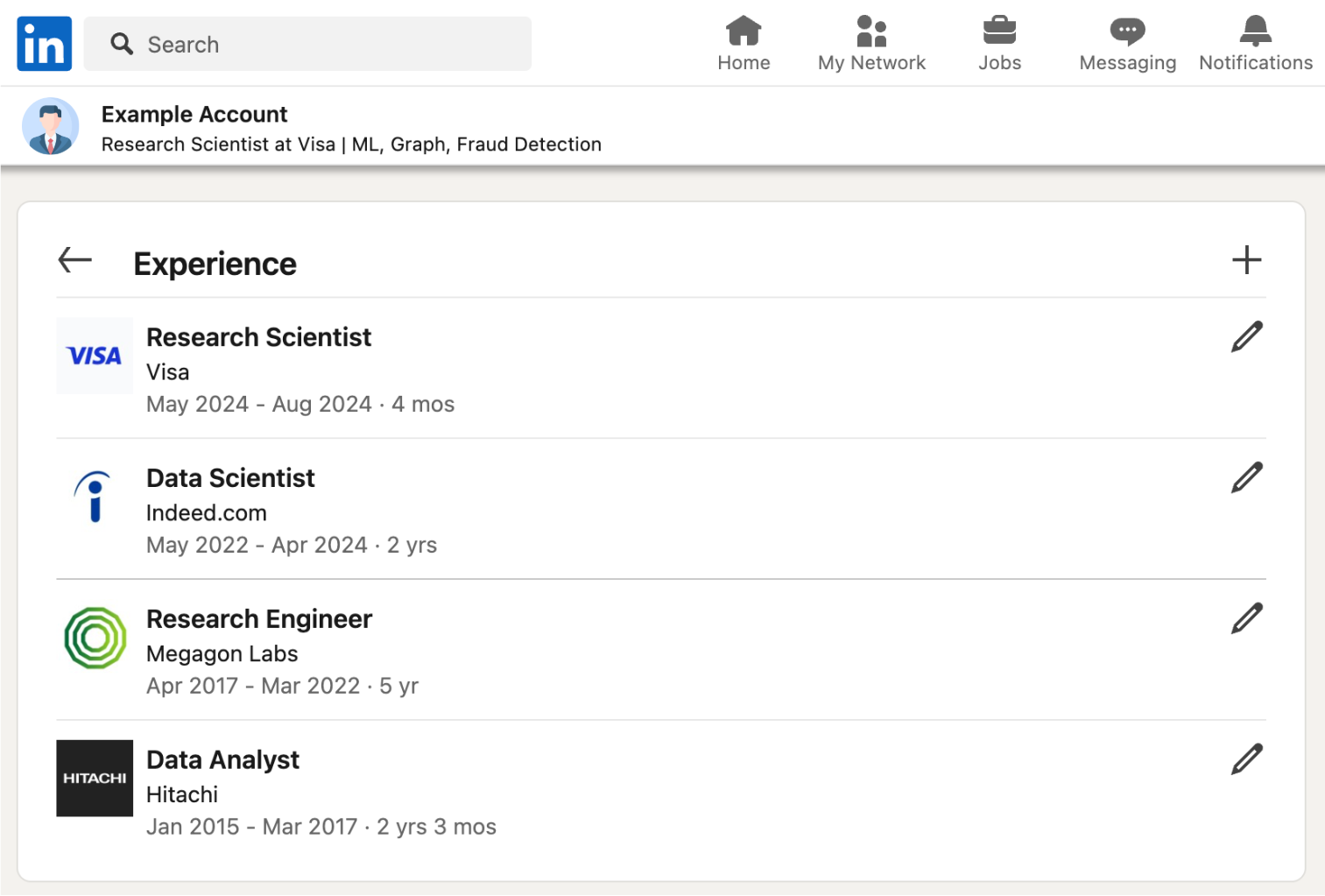}
    \caption{LinkedIn profile example.}
    \label{fig:linkedin_page}
    \vspace{10pt}
\end{figure}

\begin{figure}[th]
    \centering    
    \includegraphics[width=\linewidth]{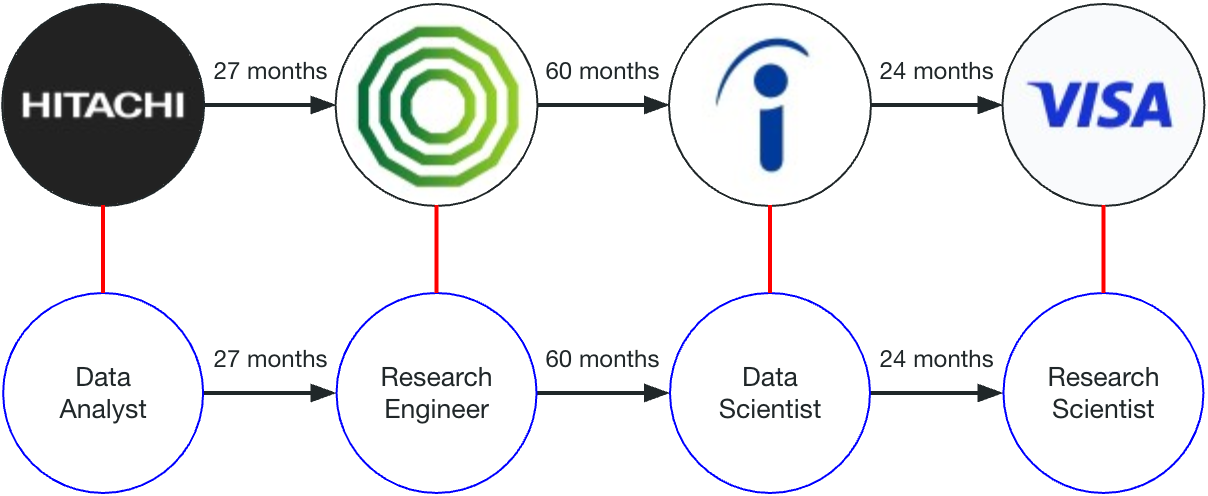}
    \caption{Graph illustration example of career trajectory.}
    \label{fig:linkedin_graph}
\end{figure}

Inspired by real-world online job platforms such as LinkedIn, we focus on core components typically emphasized in user profiles: job titles, company names, employment durations, and job descriptions. Figure \ref{fig:linkedin_page} illustrates a typical career section from a LinkedIn profile, motivating the design of our dataset. Based on this example, we can construct a corresponding career trajectory graph, as shown in \ref{fig:linkedin_graph}. 

\subsection{Illustrative Examples}
At a high level, our task aims to capture the distinction between \textit{realistic human career trajectories} and \textit{artificially generated ones}. Intuitively, genuine career paths often reflect plausible role progressions, reasonable durations, and coherent transitions across companies and positions. In contrast, machine-generated trajectories may display unrealistic or repetitive patterns that deviate from natural career development. We illustrate this distinction with intuitive examples below. 

\textbf{Genuine example:}  
\textit{Software Engineer at Stripe (14 months) $\rightarrow$ Senior SWE at Meta (28 months) $\rightarrow$ Tech Lead at Google (8 months)}  

\noindent\textbf{Synthetic example 1:}  
\textit{Software Engineer at Google (24 months) $\rightarrow$ Senior SWE at Google (36 months) $\rightarrow$ Tech Lead at Amazon (36 months) $\rightarrow$ Engineering Manager at Amazon (24 months) $\rightarrow$ Director of Engineering at Salesforce (24 months)}  

\noindent\textbf{Synthetic example 2:}  
\textit{Software Engineer at OpenAI (12 months) $\rightarrow$ Cashier at Costco (12 months) $\rightarrow$ Engineering Manager at Amazon (12 months)}  

\noindent\textbf{Synthetic example 3:}  
\textit{Software Engineer at Microsoft (9 months) $\rightarrow$ Software Engineer at Google (14 months) $\rightarrow$ Software Engineer at Microsoft (13 months) $\rightarrow$ Software Engineer at Google (14 months) $\rightarrow$ Software Engineer at Microsoft (21 months) $\rightarrow$ Software Engineer at Google (12 months)}  

These synthetic trajectories exhibit \textit{unnatural patterns}, such as implausible career transitions (e.g., SWE $\rightarrow$ Cashier $\rightarrow$ EM) or artificial timing regularities (e.g., repeated multiples of 12-month durations).

\vspace{10pt}
\section{Prompt Settings}\label{sec:prompt_settings}
\subsection{LLM-Based Generators}

\vspace{5pt}
\begin{tcolorbox}[fontupper=\small]
You are an expert career counselor generating realistic synthetic resume data for LLM benchmarking. Generate a realistic, US-based career path in the technology and business sectors. Each career path should contain at least 5 job entries (i.e., more than 4), and each entry must include the following fields:\\

- Job Title\\
- Company Name (must be a real-world company)\\
- Start Date (e.g., "Mar 2017")\\
- End Date (e.g., "May 2019")\\
- Job Duration (months) (e.g., "26")\\

Ensure each career path follows a realistic progression. Include career changes, promotions, and industry shifts where appropriate. 
\end{tcolorbox} 
\vspace{10pt}

\subsection{Agent-Based Generators}
The Agent-Based Generator is a multi-agent framework designed to produce high-quality and realistic synthetic career trajectories. It leverages an iterative conversation between two large language model (LLM) agents: the \textbf{Generator Agent} and the \textbf{Critic Agent}. Together, they simulate a human-like process of resume drafting and peer review, enabling the generation of believable and coherent career paths across the technology and business sectors. 
In our implementation, we employ two independent instances of GPT-4o, assigning one as the Generator and the other as the Critic. We select GPT-4o as the Critic given its stronger reasoning abilities in career planning and self-correction. This configuration not only improves generation quality but also reflects realistic adversarial scenarios, as malicious actors may exploit state-of-the-art LLMs such as GPT-4o to fabricate convincing resumes.

\subsubsection{System Flow}  
The generation process consists of an iterative loop involving the following steps:

\begin{enumerate}
    \item The \textbf{Generator Agent} produces an initial draft of a career trajectory, following specified structural and domain constraints.
    \item The \textbf{Critic Agent} evaluates the trajectory, identifies unrealistic elements, and provides both a realism score (1--5) and concrete suggestions for improvement.
    \item The \textbf{Generator Agent} revises the trajectory according to the Critic's feedback and produces an updated version.
    \item Steps 2 and 3 are repeated until either: the realism score reaches a threshold, or the maximum number of review rounds is reached. 
\end{enumerate}

We set {4.0} as the realism score threshold and four as the maximum number of review rounds so that Generator Agent can create career trajectories up to five times.  

\subsubsection{Job Entry Requirements} 
Each final career trajectory must contain at least 5 job entries, and each entry must include: Job Title, Company Name, Start Date, End Date, and Job Duration (months). 

This loop-based interaction reflects real-world processes of drafting and revision, enabling the Generator Agent to incorporate critical feedback iteratively. The final output is more nuanced, realistic, and diverse compared to single-pass generation methods.

\begin{tcolorbox}[fontupper=\small, colback=gray!5, colframe=gray!80!black, title=Generator Agent Prompt (Initial)]
You are an expert career counselor generating realistic synthetic resume data for LLM benchmarking. Generate a realistic, US-based career path in the technology and business sectors. Each career path should contain at least 5 job entries (i.e., more than 4), and each entry must include the following fields:\\

- Job Title\\
- Company Name (must be a real-world company)\\
- Start Date (e.g., "Mar 2017")\\
- End Date (e.g., "May 2019")\\
- Job Duration (months) (e.g., "26")\\

Ensure each career path follows a realistic progression. Include career changes, promotions, and industry shifts where appropriate.
\end{tcolorbox}

\begin{tcolorbox}[fontupper=\small, colback=gray!5, colframe=gray!80!black, title=Critic Agent Prompt]
You are a hiring manager reviewing a career trajectory. Your task is to critically evaluate the realism of the career path. \\

For each job entry, check for:\\
- Plausibility of job titles and promotions\\
- Consistency with known real-world company hiring practices\\
- Realistic time spans and durations\\
- Logical career progression and industry transitions\\

Then respond with:\\
1. A realism score from 1 (not realistic) to 5 (very realistic)\\
2. Bullet-pointed feedback on unrealistic or implausible aspects\\
3. Specific suggestions for how the Generator Agent could improve the trajectory
\end{tcolorbox}

\begin{tcolorbox}[fontupper=\small, colback=gray!5, colframe=gray!80!black, title=Generator Agent Prompt (Revision Round)]
You previously generated a career trajectory, and received the following critique:\\
\textbf{Feedback:} [Critic Feedback]\\
\textbf{Realism Score:} [Insert Score]\\

Now revise the career trajectory to address the feedback. Ensure that your updated trajectory is more realistic and aligned with real-world career progression patterns. Maintain at least 5 job entries, and include the following for each:\\

- Job Title\\
- Company Name (real-world company)\\
- Start Date\\
- End Date\\
- Job Duration (months)
\end{tcolorbox}

\subsection{LLM-Based Detectors}\label{sec:llm_detector}
\subsubsection{Zero-shot}
\begin{tcolorbox}[fontupper=\small]
You are an expert career counselor. Below is a career trajectory with multiple job entries. Your task is to determine whether this career history is likely to be real (written by a human) or synthetic (generated by an AI or a machine).\\

Please reason step-by-step considering the following aspects:\\
- Plausibility of job title transitions\\
- Realism of company-to-company moves\\
- Length and variation of job durations\\
- Presence of career gaps, promotions, or lateral shifts\\

At the end, just give a final answer: ``Real'' or ``Fake''.\\

Career Trajectory:\\
\{career\_trajectory\}\\

Answer: [Real or Fake]
\end{tcolorbox}

\subsubsection{Few-shot}
We use 4 shots as the default setting. 
\begin{tcolorbox}[fontupper=\small]
You are an expert career counselor. Your task is to determine whether a given career trajectory is real (written by a human) or fake (generated by an AI or a machine).\\

Please consider the following aspects while making your decision:\\
- Plausibility of job title transitions\\
- Realism of company-to-company moves\\
- Length and variation of job durations\\
- Presence of career gaps, promotions, or lateral shifts\\

Below are several labeled examples. Use them to guide your judgment.\\

Example 1:\\
\{example\_career\_trajectory\_1\}\\
Answer: Real\\

Example 2:\\
\{example\_career\_trajectory\_2\}\\
Answer: Fake\\
...

Example N:\\
\{example\_career\_trajectory\_N\}\\
Answer: Real\\

---\\
Now evaluate the following career trajectory using the same criteria. Just give a final answer: ``Real'' or ``Fake''\\

Career Trajectory:\\
\{career\_trajectory\}\\

Answer: [Real or Fake]
\end{tcolorbox}

\section{Realism Assessment}\label{sec:real_asses}
\begin{tcolorbox}[fontupper=\small]
You are an expert in career trajectory evaluation. 
Attached is a CSV file containing a resume. A resume consists of a sequence of job entries with columns for job\_title, company, and duration (in months). \\

Your task is to assess the realism of the resume on a scale from 1 to 5, where:\\
- 1 = Clearly artificial or unrealistic\\
- 2 = Somewhat artificial\\
- 3 = Neutral or uncertain\\
- 4 = Somewhat realistic\\
- 5 = Highly realistic and human-like\\

Please focus only on how plausible and coherent the career path sounds overall. Do not try to guess whether it was created by a human or a machine. 
\end{tcolorbox}

\section{Template for Conversion from Career Trajectory to Text}\label{sec:template}
To enable LLM-based classification, we convert structured career trajectories into a natural language format using the following template. Each job entry is represented as a sentence describing the role, company, and duration, and all entries are concatenated in reverse chronological order (from most recent to oldest):

\begin{quote}
"I have worked as a \texttt{[Job Title]} at \texttt{[Company Name]} for \texttt{[Duration]} months." 
\end{quote}

For example, a career trajectory consisting of three positions is converted as follows:

\begin{quote}
``I have worked as a Software Engineer at Google for 24 months. Before that, I was a Research Intern at OpenAI for 6 months. Prior to that, I worked as a Data Analyst at IBM for 18 months.''
\end{quote}

We apply this conversion consistently across all data samples to ensure uniformity for LLM evaluation. Additional formatting details such as joining phrases (\eg, ``Before that'', ``Prior to that'') help preserve the temporal order and improve the fluency of the resulting text.

\section{Dataset Description}\label{sec:dataset_description}
\subsection{Data Availability}
Due to the sensitive nature of career data, we only share the LLM-generated portion of our dataset, comprising 4,000 synthetic career trajectories of GPT-4o, LLaMA-3, Gemini-2.0, and Agent. 

\subsection{Dataset Structure Comparison} 
\begin{figure}[b]
    \centering
    \includegraphics[width=\linewidth]{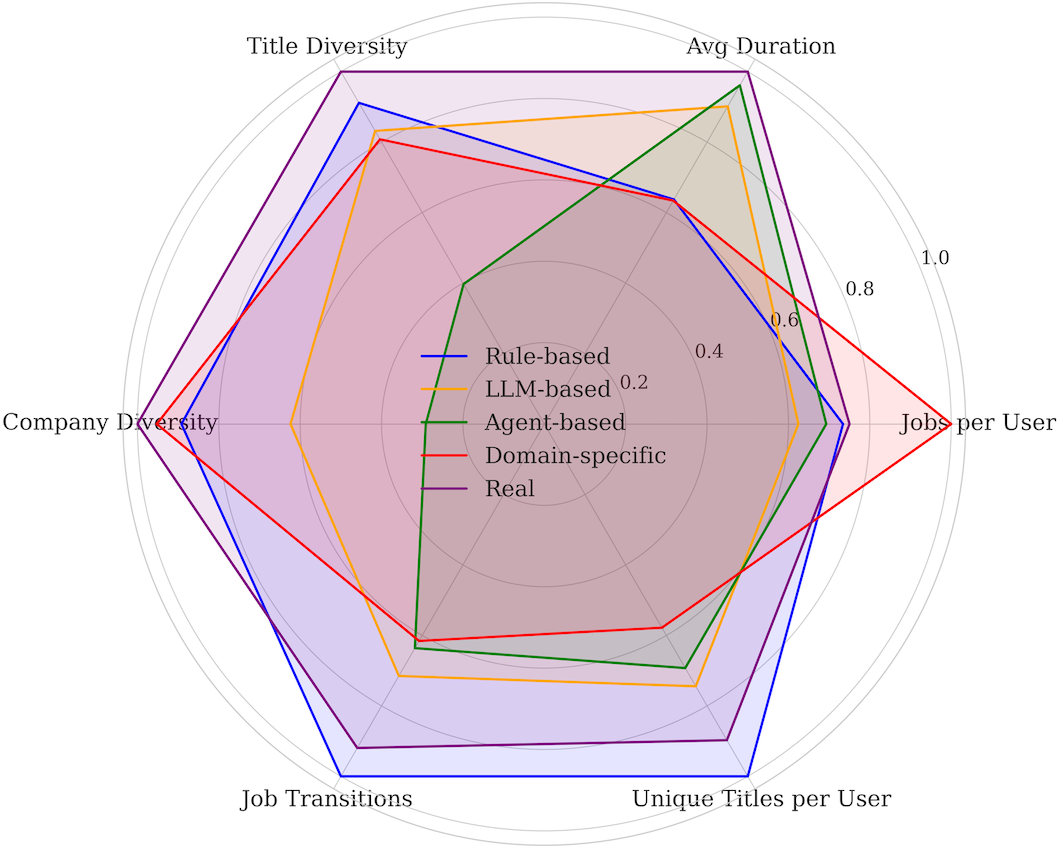}
    \caption{Generated dataset structure comparison across different generator types.}
    \label{fig:data_radar_chart} 
\end{figure}

Figure \ref{fig:data_radar_chart} illustrates the key structural characteristics across different generator types. The radar chart visualizes key structural characteristics including job density, duration patterns, diversity, and transitions. 

\end{document}